\documentclass[graybox]{svmult}

\usepackage{mathptmx, amsfonts, amsmath}       % selects Times Roman as basic font
\usepackage{helvet}         % selects Helvetica as sans-serif font
\usepackage{courier}        % selects Courier as typewriter font
\usepackage{type1cm}        % activate if the above 3 fonts are
\usepackage{makeidx}         % allows index generation
\usepackage{graphicx}        % standard LaTeX graphics tool
\usepackage{multicol}        % used for the two-column index
\usepackage[bottom]{footmisc}% places footnotes at page bottom
\usepackage{bm}

\def\N{\mathbb{N}}

\def\R{\mathbb{R}}
\let\e=\varepsilon

\let\.=\cdot
\let\0=\emptyset

\def\square{\hbox{$\sqcap\kern-7pt\sqcup$}}

\def\be{\begin{equation}}
\def\ee{\end{equation}}
\def\bea{\begin{eqnarray}}
\def\eea{\end{eqnarray}}

\makeindex             % used for the subject index
\begin{document}

\title*{Derivation of the Boltzmann equation:\\ hard spheres, short-range potentials and beyond}
% Use \titlerunning{Short Title} for an abbreviated version of
% your contribution title if the original one is too long

\author{Chiara Saffirio}
% Use \authorrunning{Short Title} for an abbreviated version of
% your contribution title if the original one is too long

\institute{Chiara Saffirio \at 
Institute of Mathematics, University of Z\"urich, Winterthurerstrasse 190, CH-8057 Z\"urich\\
\email{chiara.saffirio@math.uzh.ch}
%\and Name of Second Author \at Name, Address of Institute \email{name@email.address}
}

%
% Use the package "url.sty" to avoid
% problems with special characters
% used in your e-mail or web address
%
\maketitle

%\abstract*{ }

\abstract{We review some results concerning the derivation of the Boltzmann equation starting from the many-body classical Hamiltonian dynamics. In particular, the celebrated paper by O. E. Lanford III \cite{La75} and the more recent papers \cite{GSRT13, PSS14} are discussed.}


\section{Introduction}
\label{sec:intro}
A central question in non-equilibrium statistical mechanics is to investigate the rigorous derivation of effective macroscopic equations starting from the fundamental laws of classical mechanics. Though we are still very far from a complete understanding, considerable progress has been made in the last years in developing new mathematical methods. In particular, several interesting questions regarding classical systems in the mean-field and low-density limits are now approachable by a rigorous mathematical analysis. The aim of this paper is to give an overview on the derivation of the classical Boltzmann equation, in light of  recent developments. 

\medskip

{\bf The Boltzmann equation.} At the end of the XIXth century J. C. Maxwell (\cite{Ma67}) and L. E. Boltzmann (\cite{Bolt}) addressed independently the problem of the mathematical description of classical dilute gases, in an attempt to produce a reduced kinetic picture emerging from the microscopic fundamental laws of classical mechanics. A kinetic description holds at a mesoscopic level,
that is on quantities which averages are susceptible of measurement.
 The equation for the evolution of a rarefied gas, that nowadays bears the name of Boltzmann, reads
\be\label{eq:BE}
(\partial_t+v\cdot\nabla_x)f=Q(f,f)\,.
\ee
The unknown  $f:\R_+\times\R^3\times\R^3\to\R_+$ is the probability density of finding a particle with position $x$ and velocity $v$ at time $t$. \\
The l.h.s. in the Boltzmann Eq. (\ref{eq:BE}) is the free transport operator, representing the free flow of particles in absence of external forces. The r.h.s. is a non-linear (quadratic) operator which describes the elastic binary collisions among particles:
\be\label{eq:Q}
Q(f,f)(t,x,v)= \int_{S^2}d\omega\int_{\R^3} dv_* B(v-v_*,\omega)\{f(t,x,v')f(t,x,v_*')-f(t,x,v)f(t,x,v_*)\},
\ee
where $S^2$ is the unit sphere in $\R^3$, $\omega\in S^2$ is the  scattering vector, $v'$ and $v'_*$ are obtained as functions of $v$ and $v_*$ by the following scattering laws:
\be
\left\{
\begin{array}{l}
v'=v - [(v-v_*)\cdot\omega]\omega\,,\\
v_*'=v_*+[(v-v_*)\cdot\omega]\omega\,.
\end{array}
\right.
\ee
The integral kernel $B(\cdot,\cdot)$ is proportional to the differential cross-section.

In particular, in \cite{Bolt} Boltzmann established Eq. (\ref{eq:BE}) by taking into account the interactions among particles which occur through elastic binary collisions, which are localised in space and time. Precisely, when the particles interact as hard spheres (in other words as billiard balls), the kernel assumes the simple and explicit form $B(v-v_*,\omega)=\omega\cdot(v-v_*)$. In this case, the scattering vector $\omega$ is equal to  $\bm{\nu}\in S^2_+:=\{\theta\in S^2\ |\ \theta\cdot(v-v_*)\geq 0\}$, the unit vector pointing from the particle with velocity $v$ to the particle with velocity $v_*$. %follows: let $x$ and $x_*$ be the center of the two particles involved in the collision process, with velocity $v$ and $v_*$ respectively; we define $\bm{\nu}:=\frac{x_*-x}{|x_*-x|}\in S^2_{+}$, with $S^2_+:=\{\theta\in S^2\ |\ \theta\cdot(v-v_*)\geq 0\}$.
%\begin{figure}[htbp] 
%   \centering
%   \includegraphics[scale=0.8]{HardSpheres} 
%   \caption{The two--body scattering in the case of hard-sphere interaction.}
%   \label{fig:hs}
%\end{figure}

Then the Boltzmann collision operator for hard spheres reads
\be\label{eq:BE-HS}
Q(f,f)(t,x,v)=\int_{S^2_+} d\bm{\nu}\int_{\R^3} dv_*\,\bm{\nu}\cdot(v-v_*)\,\{f(t,x,v_*')f(t,x,v')-f(t,x,v_*)f(t,x,v)\}\,.
\ee

The peculiarity of Eq. (\ref{eq:BE}) is the following: on the one hand, it purports to describe the evolution of the density of a rarefied gas, whose dynamics is time-reversible at a microscopic level; on the other hand, the equation itself has an irreversible behaviour, with an increasing entropy (the celebrated $H$-Theorem\footnote{The $H$-Theorem asserts that the kinetic entropy  associated to the solution $f(t)$ of the Boltzmann Eq.~\eqref{eq:BE} decreases in time. More precisely, let $H(f)$ be the $H$-functional defined as the information entropy with a negative sign: \[H(f(t))=\int_{\R^3\times\R^3}dx\,dv\,f(t,x,v)[\log{f(t,x,v)}-1]\,.\] A straightforward computation shows that $H(f(t))\leq H(f(0))$.}) and trend to equilibrium.

%In \cite{Bolt} Boltzmann wrote down his equation for a system of particles interacting as hard-spheres (i.~e. billiard balls). In this case, the collision operator \eqref{eq:Q} has the following expression
%%
%\be\label{eq:BE-HS}
%Q(f,f)(t,x,v)= \int_{S^2_+}d\nu\int_{\R^3} dv_*\,\nu\cdot(v-v_*)\,\{f(t,x,v')f(t,x,v_*')-f(t,x,v)f(t,x,v_*)\}\,,
%\ee
%%  
%where $S^2_+=\{\nu\in S^2\ |\ \nu\cdot(v-v_*)\geq 0\}$...
%

\medskip
{\bf The derivation problem.} The issue of derivation consists in determining whether the theory of Boltzmann is only a phenomenological observation or a rigorous consequence of the laws of mechanics. The question is: is it possible to derive mathematically  an irreversible dynamics, such as the Boltzmann dynamics, starting from the microscopic reversible classical dynamics? A positive answer to this question would show rigorously that there is no contradiction between the reversibility of the molecular dynamics and the irreversibility implied by the $H$-Theorem.   

The mathematical formulation of the derivation problem was given by H.Grad (more than fifty years after Boltzmann). Indeed, in \cite{Gr58} Grad formulated  for the first time the question of the validity of the Boltzmann equation as a limit - involving the number of particles - in which it is expected to hold. To state Grad's idea, we introduce the key ingredients for the  description of a microscopic classical dynamics of a system of particles. 

 The time evolution of a configuration of $N$ particles in the phase space 
\be
\mathcal{M}_N=\{(q_1,\dots,q_N,v_1,\dots,v_N)\in(\R^{3}\times\R^{3})^N\ :\ |q_i-q_k|>0,\ i,k=1,\dots,N,\ k\neq i\}\,,
\ee
 is given by the Newton equations:
\be\label{eq:N}
\left\{
\begin{array}{ll}
\dot{q_i}(\tau)=v_i(\tau)\,,&\\ &\ \ \ i=1,\dots,N\\
\dot{v_i}(\tau)=-\sum_{j\neq i}^N \nabla\Phi(q_i(\tau)-q_j(\tau))\,,
\end{array}
\right.
\ee 
where $\tau$ is the time variable, $(q_1,\dots,q_N)\in\R^{3N}$ and $(v_1,\dots,v_N)\in\R^{3N}$ are respectively the position and velocity variables of the $N$ particles, and $\Phi$ is a smooth two--body interaction potential\footnote{For simplicity, the potential is assumed to be smooth, to ensure the existence and uniqueness of the solution to the Newton equations (\ref{eq:N}).}. We introduce the Hamiltonian associated to \eqref{eq:N}:
\be\label{eq:H}
H(\tau,q_1,\dots,q_N,v_1,\dots,v_N)=\frac{1}{2} \sum_{i=1}^N |v_i(\tau)|^2+ \sum_{\substack{i,j=1\\ j\neq i}}^N \Phi(q_i(\tau)-q_j(\tau))\,.
\ee
which is constant in time. 

We stress that, in order to get a kinetic picture, we are not interested in the detailed analysis of the motion of each particle, but in the collective behaviour of the system. For this reason it is useful to adopt a statistical viewpoint: consider a probability density $W_0^N$ on the phase space $\R^{3N}\times\R^{3N}$ and denote by $\Psi(\tau)$ the Newtonian flow. Then, $W^N(\tau):=W_0^N\circ\Psi(-\tau)$ solves the Liouville equation
\be\label{eq:Liouville}
\partial_\tau W^N(\tau) + \sum_{i=1}^N v_i\cdot \nabla_{q_i}W^N(\tau) -\sum_{ j\neq i}^N\nabla_{q_i}\Phi(q_i-q_j)\cdot\nabla_{v_i}W^N(\tau)=0\,
\ee
for a probability density $W^N$ on the phase space $\mathcal{M}_N$,
with $W^N\in\mathcal{C}^1(\R_+\times\mathcal{M}_N)$, $v_i\cdot\nabla_{q_i}W^N,\ \nabla\Phi\cdot\nabla_{v_i} W^N\in L^1(\mathcal{M}_N)$. Note moreover that since the particles are identical, $W^N$ is symmetric w.r.t. permutation of particles.

In \cite{Gr58}, Grad remarked that the solution of (\ref{eq:Liouville}) can be approximated by the Boltzmann equation in the following regime, and in the following sense. Let $\e>0$ be a scale parameter, which represents the ratio between macroscopic and microscopic units. Let us scale time and space according to 
\be\label{eq:scaling}
 t=\e\tau,\ \ x_i=\e q_i,\ \forall i=1,\dots,N\,.
\ee
Sometimes we will use the shortened notation $z_i:=(x_i,v_i)\in\R^3\times\R^3$, for $i=1,\dots,N$.
In the limit of $N$ large, with $N\e^2=1$, the solution of (\ref{eq:Liouville}) with approximately factorised initial data 
remains approximately factorised. Note that approximate factorisation has to be understood in the sense of the marginal distributions
\be\label{eq:marginals}
f_j^N(t,z_1,\dots,z_j):=\int_{\R^{6(N-j)}} W^N(t,z_1,\dots,z_N)\,dz_{j+1}\dots dz_{N}\,.
\ee
More precisely, if
$
f^N_j(0,x_1,\dots,x_j,v_1\dots,v_j)\simeq f_0^{\otimes j}(x_i,v_i),
$
then
\be\label{eq:prop-chaos}
f^N_j(t,x_1,\dots,x_j,v_1\dots,v_j)\simeq f(t)^{\otimes j}(x_i,v_i),
\ee
where $f$ solves the Boltzmann Eq. (\ref{eq:BE}) with initial datum $f_0$. This approximation, called {\it propagation of chaos}, is specified in Theorem \ref{thm:Lanford}.

The regime
$$N\to\infty,\ \ \ \ \ N\e^2=1$$
is called the low-density limit (or {\it Boltzmann-Grad limit}, BG-limit from now on). The underlying idea is that, on the one hand, we want to describe the physical situation in which the gas is rarefied. Hence a tagged particle undergoes a finite number of collisions in a macroscopic time, implying that the density $N\e^3$ vanishes in the limit of large $N$. On the other hand, we want the collisional structure of the microscopic system to survive in the limit, that is 
$$
\frac{\rm number\ of\ interactions}{\rm time\ unit}=O(1)\,.
$$
Therefore, it is important to keep in mind that the number of particles $N$ is linked to the scale parameter $\e$ by the relation $N\e^2=O(1)$ (for simplicity we have chosen $N\e^2=1$), so that the limit $\e\to 0$ is equivalent to  $N\to\infty$. Observe that, if $\Phi$ has compact support, the parameter $\e$ represents the range of the interaction at a macroscopic scale thanks to the scaling \eqref{eq:scaling}. 

The crucial assumption to be verified is that Eq. (\ref{eq:prop-chaos}) holds for positive times once it is assumed to be true at time zero.
The lack of correlation between two particles (i.e. factorisation (\ref{eq:prop-chaos})) up to the moment in which they collide is the way to conciliate a microscopic time-reversible dynamics with an evolution equation exhibiting increase of entropy and trend to equilibrium.

The evolution of the $j$-particle marginal distribution (\ref{eq:marginals}) is given by the following 
set of equations, called BBGKY hierarchy (after Bogoliubov \cite{Bog}, Born and Green \cite{BoGr}, Kirkwood \cite{K}, Yvon \cite{Y}). It is obtained by integrating the Liouville equation (\ref{eq:Liouville}) with respect to the variables $dz_{j+1}\dots dz_N$:
\be\label{eq:BBGKY}
\begin{split}
\partial_t  f_j^N(t,z_1,\dots,z_j)& + \sum_{i=1}^j v_i\cdot\nabla_{x_i}f^N_j(t,z_1,\dots,z_j)  \\
&= (\mathcal{L}_j^\e f_j^N)(t,z_1,\dots,z_j) + (\mathcal{C}_{j+1}^\e f_{j+1}^N)(t,z_1,\dots,z_j)\,,
\end{split}
\ee
where
\be\label{eq:liouville-op-j}
(\mathcal{L}_j^\e f_j^N)(t,z_1,\dots,z_j):= \frac{1}{\e}\sum_{i=1}^j\sum_{\substack{k=1\\k\neq i}}^j \nabla\Phi\left(\frac{x_i-x_k}{\e}\right)\cdot\nabla_{v_i}f_j^N(t,z_1,\dots,z_j)\,,
\ee
\be\label{eq:collision-op-j}
\begin{split}
(\mathcal{C}_{j+1}^\e f_{j+1}^N)&(t,z_1,\dots,z_j)\\&:= \frac{(N-j)}{\e}\sum_{i=1}^j\int\nabla\Phi\left(\frac{x_i-x_{j+1}}{\e}\right)\cdot\nabla_{v_i} f^N_{j+1}(t,z_1,\dots,z_{j+1})\,dz_{j+1}\,,
\end{split}
\ee
where %(\ref{eq:liouville-op-j}) is the interaction part of the Liouville operator  (\ref{eq:Liouville}) restricted to a set of $j$ particles, and 
(\ref{eq:collision-op-j}) is called \emph{collision operator}.

%%
%\be\label{eq:BBGKY}
%\begin{split}
%(\partial_t+\sum_{i=1}^j)f_j^N - \frac{1}{\e}\sum_{i=1}^j\sum_{\substack{k=1\\ k\neq i}}^j&\nabla\Phi\left(\frac{x_i-x_k}{\e}\right)\cdot\nabla_{v_i}f_j^N\\
%&=\frac{(N-j)}{\e}\sum{i=1}^j\int dx_{j+1}\,dv_{j+1}\nabla\Phi\left(\frac{x_i-x_{j+1}}{\e}\right)\cdot\nabla_{v_i}f_{j+1}^N\,.
%\end{split}
%\ee
%% The approximation above is specified in Theorem \ref{thm:Lanford}.
%

%

\medskip
{\bf Lanford's theorem.} The first rigorous derivation of the Boltzmann equation in the low-density limit was given by Lanford \cite{La75}, for hard-sphere potentials.
To prove his result, Lanford studied the BBGKY hierarchy, describing the evolution of the marginals $f_j^N$, $j=1,\dots N$, and expressed   
$f_j^N(t)$ as a sum of operators acting on the sequence of initial data $f_j^N(0)$. To state in a precise way the result, we need to introduce some functional normed spaces, on which these operators act. 
\begin{definition}\label{def:fspace}
Let $X_{j,\beta}$ be the space of Borel functions $f_j$ on $\mathcal{M}_{j}$ such that
\[ \|f_j\|_{j,\beta}=\sup_{(z_1,\dots,z_j)\in\mathcal{M}_j} |f_j(z_1,\dots,z_j)|(\beta/2\pi)^{-\frac{3}{2}j}\exp({\beta H(z_1,\dots,z_j)}) <\infty\,, \]
where $H(z_1,\dots,z_j)$ is the Hamiltonian of the $j$ particle system.
\end{definition} 
%
%Notice that the absolute value of elements of the functional space $X_{j,\beta}$  is point-wise bounded from above by $\exp{[-\beta\sum_{i=1}^j \frac{v_i^2}{2}]}$. 

\begin{definition}\label{def:fjspace}
For all $b>0$, we define the space $X_{b,\beta}$ of sequences of functions $f=\{f_j\}_{j\geq 1}$ such that the following norm is finite 
\[  \|f\|_{b,\beta}=\sup_{j}\ b^{-j}\|f_j\|_{j,\beta}\,.  \]
\end{definition}

\begin{remark}
Observe that $\beta$ and $b$ can be interpreted respectively as the inverse of the temperature and the activity of the $j$-particle system, see \cite{Ruelle}. \end{remark}

Now we have all the ingredients to state Lanford's  theorem: 

\begin{theorem}[Lanford 1975]\label{thm:Lanford}
Given a system of $N$ identical hard spheres of diameter $\e$ and the set $f^N=\{f_j^N\}_{1 \leq j\leq N}$ of associated $j$-particle marginals. Assume that: 
\begin{itemize}
\item[(i)] \ there exist positive constants $b$ and $\beta$ such that 
\[\|f^N(0)\|_{b,\beta}\leq C\,,\] 
where $C$ is an absolute constant, independent of $N$;
\item[(ii)] \ $f_j$ is continuous on the phase space $\mathcal{M}_j$ and 
\[ \lim_{\e\to 0} f^N_j(0,x_1,\dots,x_j,v_1,\dots,v_j) = f(0)^{\otimes j}(x,v) \]
uniformly on compact sets in $\mathcal{M}_j$. 
\end{itemize}
Then, there exists a strictly positive time $t_0:=[K\,\pi\,N\e^2 b\,\beta^{-1/2}]^{-1}$, with $K$ a positive constant, such that \ for $0 < t < t_0$,  $$f^N_j(t,x_1,\dots,x_j,v_1,\dots,v_j)\to  f(t)^{\otimes j}(x,v)$$
 a.e. in the BG-limit, with $f(t,\cdot,\cdot)$ the solution of the Boltzmann equation with initial datum $f(0,\cdot,\cdot)$.
\end{theorem}

\begin{remark}
Notice that in the hard-sphere case, the system \eqref{eq:N} is defined for a singular potential on the phase space 
\begin{equation}\label{eq:M-eps}
\mathcal{M}_N(\e):=\{ (x_1,\dots,x_N,v_1,\dots,v_N)\in\R^{3N}\times\R^{3N}\ :\ |x_i-x_j|\geq\e,\ \mbox{for}\ i\neq j \}
\end{equation} 
and the $j$-particle marginals are defined accordingly to this constraint.
\end{remark}

\begin{remark}
Observe that the Boltzmann equation only describes likely configurations, i.e. there are configurations which are out of the picture painted by Boltzmann. This justifies the almost everywhere convergence.  
\end{remark}

Although all the ideas of the proof were present in \cite{La75}, some details were missing and they have been analysed  in \cite{Sp84, Sp85, Uc88, CIP94, GePe, Uk01, GSRT13, PSS14}.
 We will give a sketch of the proof of Theorem \ref{thm:Lanford} in Sect.~\ref{sect:HS}. A slightly different but detailed argument can be found in \cite{GSRT13}, Part II.

We stress that Lanford's result holds only for short time intervals, which are of the same order as the mean free time. This is a severe limitation, since in the applications of the Boltzmann equation a long-time behaviour of the solution is involved. One of the difficulties in extending the proof for long times is to prove that, once the $j$-particle marginals $f_j^N(0)$ are smooth, their evolutions $f_j^N(t)$ do not develop singularities. To our knowledge, the only situation in which the validity result for the non linear Boltzmann equation has been proved globally in time is the one analysed in \cite{IP1,IP2}, where a rare cloud of gas expanding in the vacuum is considered. Nevertheless, the positive time $t_0$ in Lanford's theorem is large enough to observe a decrease of entropy in the Boltzmann $H$-functional. It is worth mentioning that recently a new quantitative point of view to study the correlations has been introduced in \cite{PS14}, where the authors consider a system of hard spheres in the BG-limit and introduce a set of functions measuring the correlation error. Although these  objects seem to be more appropriate to identify and isolate the dynamical events responsible for the breakdown of propagation of chaos, the extension to long times of the validity of the Boltzmann equation is still far from being achieved. 
Recently, the validity for long times has been achieved in \cite{BGSR14} and \cite{BGSR15}, in the context of the linear Boltzmann equation in any dimension $d\geq 2$ and the linearised Boltzmann equation in dimension $d=2$.  
\medskip 

The second limitation of Lanford's theorem is the restriction to the hard-sphere interaction. In 1975 King presented a PhD thesis (unpublished, \cite{Ki75}) on the derivation of Eq. (\ref{eq:BE}) for smooth short-range potentials. This problem has been considered a simple extension of Lanford proof, until it was recently reconsidered in \cite{GSRT13}: there the authors proved rigorously that Eq. (\ref{eq:BE}) can be obtained from a system of particles interacting via  smooth positive short-range potentials. This is done through a sophisticated analysis and some further restrictions on the potential are needed. Hence, in \cite{GSRT13} the authors have shown that the extension from hard-spheres to short-range potentials is a delicate and non trivial task.

The paper is organised as follows: in Sect.~\ref{sect:HS} we give a sketch of the proof of Theorem \ref{thm:Lanford}, where the derivation of the Boltzmann equation for the hard-sphere dynamics is presented; in Sect.~\ref{sect:short} we give an idea of the main difficulties in extending Theorem \ref{thm:Lanford} to the case of short-range potentials and we  review the recent results obtained in \cite{GSRT13, PSS14}; Sect.~\ref{sect:long} is devoted to the open problem of the derivation of Eq. (\ref{eq:BE}) in the case of long-range interactions.   

%%%%%%%%%%%%%%%%%%%%%%%%%%%%%%%%%%%%%%%%%%%%%
%%%%%%%%%%%%%%%%%%%%%%%%%%%%%%%%%%%%%%%%%%%%%

\section{Hard-sphere interaction}
\label{sect:HS}

The aim of this section is to give an overview on the steps of Lanford's proof \cite{La75}. 

We consider a system of $N$ particles, interacting as hard spheres of diameter  $\e$ on the phase space \eqref{eq:M-eps}; we define the $j$-particle marginals associated to it and we compute their evolution in time, according to \eqref{eq:BBGKY}.
An important observation is that, as already pointed out in \cite{Ce72}, there is a formal similarity between the BBGKY hierarchy for hard spheres and the Boltzmann equation. Indeed, in the case of a hard-sphere interaction, the BBGKY
 reads as  \eqref{eq:BBGKY}, with $\mathcal{L}_j^\e$ replaced by appropriate boundary conditions $\tilde{\mathcal{L}^\e_j}$ and $C_{j+1}^\e$ is replaced by
\be \begin{split}
&(\tilde{\mathcal{C}}_{j+1}^\e f_{j+1}^N)(t,z_1,\dots,z_j)=(N-j)\e^2\sum_{i=1}^j\int d\omega\int dv_{j+1}\, \omega\cdot(v_{j+1}-v_i)\\
&\ \ \ \ \ \ \ \ \ \ \ \ \ \ \ \ \ \ \ \ \ \ \ \ \ \ \ \ \ \ \ \ \ \ \ \ \ \ \ \ \ \ \ \ \ \ \ \ \ \ \times f_{j+1}^N(t,x_1,\dots,x_j,x_i+\e\omega,v_1,\dots,v_{j+1})\,.
\end{split}
\ee
To deal with the full differential hierarchy is a hard task. Indeed one has to deal with a family of $N$ integro-differential equations, in the limit of $N$ large. Hence, the idea of Lanford is to proceed in a perturbative way, by considering the temporal series solution of the BBGKY hierarchy, i.~e. the Duhamel series
\be\label{eq:seriesHS}
\begin{split}
f_{j}^N(t,z_1,\dots,z_j)=\sum_{n= 0}^{N-j} \alpha_n^\e(j)\int_0^t dt_1\int_0^{t_1} dt_2\dots&\int_0^{t_{n-1}} dt_n \mathcal{S}_j^\e(t-t_1)\tilde{\mathcal{C}}_{j+1}^\e\mathcal{S}_{j+1}^\e(t_1-t_2)\\
&\dots\tilde{\mathcal{C}}_{j+n}^\e\mathcal{S}_{j+n}^\e(t_n)f^N_{j+n}(0,z_1,\dots,z_{j+n})\,,
\end{split}
\ee
where $\alpha_n^\e(j):=\e^{2n}(N-j)(N-j-1)\dots(N-j-n+1)$ is $O(1)$ in the BG-limit  and $\mathcal{S}_j^\e(t)$ is the flow operator of the $j$-body hard-sphere dynamics. Roughly speaking, it behaves as the free flow up to the first impact time, then a collision occurs according to the scattering and the dynamics restarts as a free flow with the new outgoing configuration as initial condition up to the next impact time. Notice that, by conservation of energy, the operator $\mathcal{S}_j^\e(t)$ acts as a one-parameter group of isometries on the functional space $X_{j,\beta}$, i.~e.   $\|\mathcal{S}_j^\e(t)f_j\|_{j,\beta}=\|f_j\|_{j,\beta}$ for any $\beta$. 

We want to compare \eqref{eq:seriesHS} with $f_j(t,z_1,\dots,z_j)$, obtained as follows: let $f(t,z)$ be a solution to the Boltzmann equation, then 
\[f_j(t,z_1,\dots,z_j):=\prod_{i=1}^j f(t,z_i)\]
is a solution to the following hierarchy of equations  
\be
\partial_t f_j +\sum_{i=1}^j v_i\cdot \nabla_{x_i}f_j=\mathcal{C}_{j+1}f_{j+1}
\ee
where 
\begin{equation}\label{eq:coll} \mathcal{C}_{j+1} =\sum_{k=1}^j \mathcal{C}_{k,j+1},\ \ \  \mathcal{C}_{k,j+1} = \mathcal{C}^{+}_{k,j+1} - \mathcal{C}^{-}_{k,j+1} \end{equation}

\[
\begin{split}
\mathcal{C}^{+}_{k,j+1}f_{j+1}&(t,z_1,\dots,z_j) \\=& \int_{S^2_+} d\omega\int_{\R^3} dv_{j+1}
\omega\cdot(v_k-v_{j+1}) f_{j+1}(t,z_1,\cdots,x_k,v'_k,\cdots,z_j,x_k,v'_{j+1})\,,\\
 \mathcal{C}^{-}_{k,j+1}f_{j+1}&(t,z_1,\dots,z_j) \\=& \int_{S^2_+} d\omega\int_{\R^3} dv_{j+1}
\omega\cdot(v_k-v_{j+1}) f_{j+1}(t,z_1,\cdots,x_k,v_k,\cdots,z_j,x_k,v_{j+1})\;.
\end{split}
\]
Again, we can apply iteratively Duhamel formula to get the series expansion
\be
\begin{split}\label{eq:seriesBE}
f_j(t,z_1,\dots,z_j)=\sum_{n\geq 0} \int_0^t dt_1\int_0^{t_1} dt_2\dots\int_0^{t_{n-1}} &dt_n \mathcal{S}_j(t-t_1)\mathcal{C}_{j+1}\mathcal{S}_{j+1}(t_1-t_2)\\
&\dots\mathcal{C}_{j+n}\mathcal{S}_{j+n}(t_n)f_{j+n}(0,z_1,\dots,z_{j+n})\,,
\end{split}
\ee
where $\mathcal{S}_j(t)$ is the free-flow of $j$ particles.

Lanford's proof is made of two parts: 
\begin{itemize}
\item[$(a)$]\ a proof of the absolute convergence of series expansions \eqref{eq:seriesHS} and \eqref{eq:seriesBE}, uniformly in $\e$; 
\item[$(b)$]\ a proof of the term by term convergence of one series to the other in the BG-limit. 
\end{itemize}
The limitation to small times arises from point $(a)$. Indeed, we prove the absolute convergence of the series by bounding the series by the geometric series $\sum_n (Ct)^n$, for which the convergence is achieved only when $|t|<1/C$. 
\medskip

{\bf Step $(a)$: absolute convergence of the series.}
In the first step we show that the series solutions exist, at least in a small time interval, by proving the absolute convergence of \eqref{eq:seriesHS} and \eqref{eq:seriesBE}. 
We focus on Eq. \eqref{eq:seriesHS}, the procedure for Eq. \eqref{eq:seriesBE} is analogous. 

We observe that Eq. \eqref{eq:seriesHS} expresses $f_j^N(t)$ as a sum of operators acting on the sequence of initial data $f_j^N(0)$, hence it is useful to set the problem on the functional spaces introduced in Definitions \ref{def:fspace} and \ref{def:fjspace}, on which the operators $\mathcal{S}^\e_j$ and $\tilde{\mathcal{C}}_{j+1}^\e$ act. 

To prove the absolute convergence of the series we first prove the following 
\begin{proposition}\label{prop:main-est}
Let $\beta>\beta'>0$ and $b'>(\beta'/\beta)^{3/2}b$. Then, for all $f^N=\{f_j^N\}_{j\geq 1}\in X_{b,\beta}$, there exists a constant $K=K(\beta'/\beta,b'/b)$ such that 
\be\label{eq:main-est}
%\begin{split}
\sup_j b'^{-j}\|\mathcal{S}_j^\e(t-t_1)\tilde{\mathcal{C}}^\e_{j+1}\mathcal{S}^\e_{j+1}(t_1-t_2)\dots\tilde{\mathcal{C}}^\e_{j+n}\mathcal{S}^\e_{j+n}(t_n)f_{j+n}^N(0)\|_{j,\beta'}
\leq n!\,t_0^{-n} \|f^N\|_{b,\beta}\,,
%\end{split}
\ee
where $t_0=[K\pi N\e^2 b \beta^{-1/2}]^{-1}$.

\end{proposition}

\begin{remark}
The time of validity $t_0$ is of the order of the mean-free time, defined as the ratio between the mean-free path and the mean square velocity.
\end{remark}
\begin{remark}
Observe that we started from a functional space $X_{b,\beta}$ and we obtained $X_{b',\beta'}$. The loss is quantised by $\beta-\beta'$ and $b-b'$ and it will be compensated by integration in time (see Corollary \ref{cor:est-series}). This is typical of Cauchy-Kovaleskaya proofs (see also \cite{Uk01}).
\end{remark}

\begin{proof} 
We first estimate the term $\|\tilde{\mathcal{C}}_{j+1}^\e f_{j+1}^N\|_{b',\beta'}$. We have 
\[|f_{j+1}^N(z_1,\dots,z_{j+1})|\leq\|f_{j+1}^N\|_{j+1,\beta}(\beta/2\pi)^{\frac{3}{2}j}e^{-\beta\sum_{i=1}^{j+1} \frac{v_i^2}{2}}\,.\]
Hence,
\[
\begin{split}
|\tilde{\mathcal{C}}_{j+1}^\e &f_{j+1}^N(z_1,\dots,z_j)| \\
\leq & \pi N\e^2 \|f_{j+1}^N\|_{j+1,\beta} \int dv_{j+1}\,\sum_{i=1}^j(|v_i|+|v_{j+1}|)\left(\frac{\beta}{2\pi}\right)^{\frac{3}{2}j}e^{-\beta\sum_{i=1}^{j+1}\frac{v_i^2}{2}}\,.
\end{split}
\]
Therefore, simple computations show that 
\be\label{eq:C-est}
\|\tilde{\mathcal{C}}_{j+1}^\e f_{j+1}^N\|_{j,\beta'}\leq \pi N\e^2 \left(\frac{\beta}{\beta'}\right)^{\frac{3}{2}j}\left[(4\pi/\beta)^{3/2}\frac{\sqrt j}{\sqrt{\beta-\beta'}}+j\frac{8\pi}{\beta^2}\right]\,\|f_{j+1}^N\|_{j+1,\beta}\,
\ee
so that $\tilde{\mathcal{C}}_{j+1}^\e$ is a bounded operator from $X_{j+1,\beta}$ to $X_{j,\beta'}$ for any $\beta>\beta'$.
 The bound we got depends on $j$ as $j(\beta/\beta')^{\frac{3}{2}j}$, hence, for  $b'>(\beta/\beta')^{3/2}b$,  the sequence of operators $\{\tilde{\mathcal{C}}_{j+1}^\e\}_{j\geq 1}$ is a bounded operator from $X_{b,\beta}$ to $X_{b',\beta'}$.

Since $\mathcal{S}_j^\e(t)$ is an isometry on $X_{j,\beta}$, by iterating the argument above, we obtain the bound \eqref{eq:main-est}.
\end{proof}

\begin{corollary}\label{cor:est-series}
Let $b,b',\beta,\beta',t_0$ given as in Proposition \ref{prop:main-est}. Then the $n$-th term in \eqref{eq:seriesHS} 
\[
T^\e(j,n)=\mathcal{S}_j^\e(t-t_1)\tilde{\mathcal{C}}_{j+1}^\e\mathcal{S}_{j+1}^\e(t_1-t_2)\dots\tilde{\mathcal{C}}_{j+n}^\e\mathcal{S}_{j+n}^\e(t_n)
\]
is an operator $T^\e(j,n):X_{b,\beta}\to X_{b',\beta'}$, such that the following bound holds  
\be\label{eq:cor}
\|T^\e(j,n)f^N\|_{b,\beta}\leq C\left(\frac{|t|}{t_0}\right)^n\,. 
\ee
Therefore the series \eqref{eq:seriesHS} converges uniformly in $\e$, $N$ and $j$, for $|t|<t_0$.
\end{corollary}
\begin{remark}
Observe that we started from a functional space $X_{b,\beta}$ and we obtained $X_{b',\beta'}$. The loss is quantised by $\beta-\beta'$ and $b-b'$ and it will be compensated by integration in time (see Corollary \ref{cor:est-series}). This is typical of Cauchy-Kovaleskaya proofs (see also \cite{Uk01}). Due to the singularity of the interaction, in the hard-sphere case this is a delicate argument that has been made rigours in  \cite{S} and in the erratum of \cite{GSRT13}.
\end{remark}

\begin{proof}
The time integrals can be easily bounded as follows
\[\int_0^t dt_1\int_0^{t_1} dt_2\dots\int_0^{t_{n-1}}dt_n\leq \frac{t^n}{n!}\,.\]
Observe that in the BG-limit $\alpha_n^\e(j)=O(1)$ uniformly in $j$. By virtue of Proposition~\ref{prop:main-est}, estimate \eqref{eq:cor} follows.
\end{proof}

\medskip

{\bf Step $(b)$: term by term convergence.}
To prove the convergence of each term of \eqref{eq:seriesHS} to the corresponding term in \eqref{eq:seriesBE}, we need to look at the structure of each term of the series expansion. In order to have a clearer picture, it is useful to rewrite first \eqref{eq:seriesBE} in a handier way, which expresses the terms of the series through binary trees. 
The analysis which follows is called {\it tree expansion} and it relays strongly on an interpretational effort while considering \eqref{eq:seriesBE} (or \eqref{eq:seriesHS}). Basically, we look at \eqref{eq:seriesBE} and we consider the $j$ particles $(z_1,\dots,z_j)$ to have known positions and velocities. In the r.h.s., the integrand describes a collision process, in which a particle $j+1$ is added to the $j$ fixed particles via the definition \eqref{eq:coll} of collision operator. The tree expansion is based on this interpretation of the iteration of transport flow and collision process. 
Precisely, for each $j$ and $n$, we denote by $\Gamma(j,n)$ the binary tree with $j$ roots and $n$ nodes. For fixed $j$ and $n$, each tree $\Gamma(j,n)$ represents a class of backwards trajectories 
\[  \zeta(s)=(\xi(s),\eta(s))\,,\ \ \ s\in(0,t)\,, \]
called the Boltzmann backwards flow (BBF), and specified by  the collection of variables in the r.h.s. of \eqref{eq:seriesHS}. the $j$-particle configuration at time $t$ is denoted by $(z_1,\dots,z_j)$; $n$ is the number of added particles; $t_1,\dots,t_n$ are the times of creation of the added particles; $\bm{\nu}_1,\dots,\bm{\nu}_n$ are the impact vectors of the added particles; $v_{j+1},\dots,v_{j+n}$ are the velocities of the added particles; $\sigma_1,\dots,\sigma_{n}$ indicate the type of creation, i.~e. outgoing when $\sigma_{i}=+$ or incoming when $\sigma_i=-$. %(see \cite{PSS14}).

Then the Boltzmann series \eqref{eq:seriesBE} can be rewritten as
\be\label{eq:treeBE}
f_j(t,z_1,\dots,z_j)=\sum_{n\geq 0} \mathcal{T}(j,n)\,,
\ee
with
\[\mathcal{T}(j,n):=\sum_{\Gamma(j,n)} \prod_{i=1}^n \sigma_i \int d\Lambda\,\left(\prod_{i=1}^n B_i\right)\,f_{j+n}(0,\zeta(0))\,, \]
where 
\[ d\Lambda= \bm{1}_{\{t_1>t_2>\dots>t_n\}} dt_1\dots dt_n\,d\bm\nu_1\dots d\bm\nu_n\,dv_{j+1}\dots dv_{j+n} \]
and for $i=1,\dots,n$
\[ B_i=|\bm\nu_i\cdot(v_{j+i}-\eta_{k_i}(t_i^+))|\bm{1}_{\{\sigma_i\bm\nu_i\cdot(v_{j+i}-\eta_{k_i}(t_i^+))\geq 0\}}\,,  \]
with $k_i$ the index of the progenitor of particle $j+i$ in the binary tree (see \cite{PSS14}).

Analogously, Eq. \eqref{eq:seriesHS} can be rewritten as
\be\label{eq:treeHS}
f_j^N(t,z_1,\dots,z_j)= \sum_{n=0}^{N-j}\alpha_{n}^\e(j)\mathcal{T}^\e(j,n)\,,
\ee
with
\[\mathcal{T}^\e(j,n):=\sum_{\Gamma(j,n)}\prod_{i=1}^n \sigma_i\int{d\Lambda}\,\left(\prod_{i=1}^n B_i^\e\right)\,f_{j+n}^N(0,\zeta^\e(0))\,,\]
where $\zeta^\e(s)=(\xi^\e(s),\eta^\e(s))$ is a backward in time flow associated to the particle dynamics, called the interacting backwards flow (IBF); the integral kernel is given by
\[B_i^\e=|\bm\nu_i\cdot(v_{j+1}-\eta_{k_i}(t_i^+))|\bm{1}_{\{ \sigma_i\,\bm\nu_{i}\cdot(v_{j+1}-\eta_{k_i}(t_i^+))\geq 0 \}}\bm{1}_{\{ |\xi^\e_{j+i}(t_i)-\xi^\e_k(t_i)|>\e,\ \forall\,k\neq k_{i} \}}\,,\]
with $k_i$ the index of the progenitor of particle $i$ in the binary tree.
That is, the IBF $\zeta^\e$ is constructed analogously to the BBF $\zeta$ with the difference that, between two creations, the trajectories evolve according to the interaction operator $\mathcal{S}^\e$. Moreover, the created particles are added at distance $\e$ from their progenitors in the tree. 

By means of this expansion, the proof of step $(b)$ reduces, via dominated convergence arguments, to the proof of a.e. convergence of the IBF to the BBF:
\be\label{eq:flow-convergence}
\zeta^\e(s)\to\zeta(s)\,,\ \ \ \mbox{a.e. with respect to $d\Lambda$, for every $s\in(0,t)$} 
\ee
where  $t<t_0$ is given, with $t_0$ the limiting time obtained in step $(a)$. 

We on the generic terms of the two series \eqref{eq:treeHS} and \eqref{eq:treeBE}, for fixed $j$ and $n$, we consider the difference
 \be\label{eq:difference}
 |\alpha_n^\e(j)\mathcal{T}^\e(j,n)-\mathcal{T}(j,n)|\leq |(\alpha_n^\e(j)-1)\mathcal{T}^\e(j,n)| + |\mathcal{T}^\e(j,n)-\mathcal{T}(j,n)|\,.
 \ee
Since $\alpha_n^\e(j)\to 1$ in the BG-limit, the first term in the r.h.s. of Eq. \eqref{eq:difference} vanishes as $N\to\infty$ and $N\e^2=O(1)$. As for the second term in the r.h.s. of Eq. \eqref{eq:difference}, we split it as follows:
\be\label{eq:split}
\begin{split}
|\mathcal{T}^\e(j,n)&-\mathcal{T}(j,n)|
\\&\leq |\sum_{\Gamma(j,n)}\sum_{\sigma_1,\dots,\sigma_n}(-1)^{|\bm\sigma|}\int{d\Lambda}\,[\prod_{i=1}^n B_i^\e-\prod_{i=1}^n B_i]\,f_{j+n}^N(0,\zeta^\e(0))|\\
&+| \sum_{\Gamma(j,n)}\sum_{\sigma_1,\dots,\sigma_n}(-1)^{|\bm\sigma|}\int{d\Lambda}\,\prod_{i=1}^n B_i\,[f_{j+n}^N(0,\zeta^\e(0)) - f_{j+n}(0,\zeta^\e(0)) ] |\\
&+| \sum_{\Gamma(j,n)}\sum_{\sigma_1,\dots,\sigma_n}(-1)^{|\bm\sigma|}\int{d\Lambda}\,\prod_{i=1}^n B_i\,[f_{j+n}(0,\zeta^\e(0)) - f_{j+n}(0,\zeta(0)) ] |\,.
\end{split}
\ee
The second term on the r.h.s. of Eq. \eqref{eq:split} vanishes thanks to hypothesis $(ii)$ in Theorem \ref{thm:Lanford}; the first and the third term vanish by continuity and dominated convergence once we assume \eqref{eq:flow-convergence}.  

Hence, it remains to prove Eq. \eqref{eq:flow-convergence}. We observe that, by construction, the IBF $\zeta^\e$ differs from the BBF $\zeta$ because
 from the one hand the particle flow is sensitive to small perturbations, so that a small variation of velocities may prevent a collision, producing a drastically different flow; on the other hand in the IBF two particles may undergo a recollision, that is a collision which is not a creation (i.~e. a node of the binary tree), while in the BBF all the scattering events are creations of a new particle.
The final argument in the proof consists in the verification that the set of integrated variables such that one of the two situations above occurs, has measure zero with respect to $d\Lambda$.  The control of the recollision set is a  delicate task and one has to do an accurate analysis of the recollision set. Observe that the recollision set is a non-countable union of zero-measure sets for $\e>0$ (\cite{GePe}). The proof is concluded by showing that the set of velocities and impact vectors of the particles added in the binary tree which lead to a recollision has a vanishing measure in the BG-limit.

% Always give a unique label
% and use \ref{<label>} for cross-references
% and \cite{<label>} for bibliographic references
% use \sectionmark{}
% to alter or adjust the section heading in the running head
%%%%%%%%%%%%%%%%%%%%%%%%%%%%%%%%%%%%%%%%%%%%%
%%%%%%%%%%%%%%%%%%%%%%%%%%%%%%%%%%%%%%%%%%%%%

\section{Short-range interactions}
\label{sect:short}

In this section we present the recent results obtained in \cite{GSRT13} and \cite{PSS14}. These papers rely strongly on the ideas presented in \cite{Ki75, La75} and make use of the reduced marginals introduced by Grad. For this propose, we revert to general interaction potentials $\Phi$, with the property of being compactly supported. 

The new difficulties one has to face are essentially three: the long time scattering, the multiple collisions and the recollisions, which require a more careful analysis with respect to the hard-sphere case addressed by Lanford.  Moreover, the appropriate objects to study are not the particle marginals, but the so called {\it reduced particle marginals}, as already noted in \cite{Ki75}. The notion of reduced marginal was introduced by Grad in \cite{Gr58} and it is asymptotically (in the BG-limit) equivalent to the one of marginal. The $j$-particle reduced marginal is defined as follows:
\be\label{eq:reduced-marg}
\tilde{f}_j^N(z_1,\dots,z_j):=\int_{S(x_1,\dots,x_j)^{N-j}} W^N(z_1,\dots,z_N)\,dz_{j+1}\dots dz_N\,,
\ee
where $S(x_1,\dots,x_j)=\{(x,v)\in\R^3\times\R^3\ :\ |x-x_i|>\e,\ \mbox{for all } i=1,\dots,j\}$. The evolution equations for the reduced marginals $\tilde{f}_j^N$ are obtained by integrating the Liouville Eq. \eqref{eq:Liouville} on the domain $S(x_1,\dots,x_j)^{N-j}$ with respect to $dz_{j+1}\dots dz_N$. This procedure leads to the following hierarchy:
\be\label{eq:BBGKY-shortpot}
(\partial_t+\sum_{i=1}^j v_i\cdot\nabla_{x_i})\tilde{f}_j^N(t)=\mathcal{L}_j^\e\tilde{f}_j^N+\sum_{m=0}^{N-j-1}\mathcal{A}_{j+1+m}^\e\tilde{f}_{j+1+m}^N\,,
\ee
with
\[
\begin{split}
\mathcal{A}_{j+1+m}^\e&\tilde{f}_{j+1+m}^N(t)\\
&=\alpha_{m+1}^\e(j)\sum_{i=1}^j\e^2 \int_{S^2}d\bm\nu\bm{1}_{\{\min_{l=1,\dots,j; l\neq i}|x_i+\bm\nu\e-x_l|>\e\}}(\bm\nu)\int_{\R^3} dv_{j+1}\,\bm\nu\cdot(v_{j+1}-v_i)\\
&\times\int_{\Delta_m(x_{j+1})}\e^{-2m}\frac{dz_{j+2}\dots dz_m}{m!}\tilde{f}_{j+1+m}^N(t,z_1,\dots,z_j,x_i+\e\bm\nu,v_{j+1},z_{j+2},\dots,z_{m})
\end{split}
\]
where $\alpha_m^\e(j)=\e^{2m}(N-j)(N-j-1)(\dots)(N-j-m+1)$ and
\[
\begin{split}
\Delta_m(x_1,\dots,x_{j},x_i&+\bm{\nu}\e):=\\&\{(z_{j+2},\dots,z_m)\in S(x_1,\dots,x_j)^m\ :\ \forall\ l=j+2,\dots,j+1+m,\\
& \exists\ h_1,\dots,h_r\in\{j+2,\dots,j+1+m\} \mbox{such that } |x_l-x_{h_1}|\leq\e,\\ 
&|x_{h_{k-1}}-x_{h_k}|\leq\e,\ \mbox{for } k=2,\dots,r\ \mbox{and } \min_{i\in\{l,h_1,\dots,h_r\}}|x_i-x_{j+1}|\leq\e \}\,.
\end{split}
\]
In particular, for $m=0$
\[
\begin{split}
\mathcal{A}_{j+1}^\e\tilde{f}_{j+1}^N&=\e^2(N-j)\sum_{i=1}^j\int_{S^2}d\bm\nu\int_{\R^3}dv_{j+1}\,\bm{1}_{\{\min_{l=1,\dots,j;\ l\neq i}|x_i+\e\bm\nu-x_l|>\e\}}(\bm\nu)\\
&\times\bm\nu\cdot(v_{j+1}-v_i)\tilde{f}_{j+1}^N(t,z_1,\dots,z_j,x_i+\e\bm\nu,v_{j+1})\\
&=\e^2(N-j)C_{j+1}^\e\tilde{f}_{j+1}^N(t,z_1,\dots,z_j)\,.
\end{split}
\]
It is not difficult to prove that the contributions given by $m\geq 1$ (corresponding to multiple collisions) are negligible in the BG-limit (indeed, clearly $|\Delta_m|\sim O(\e^{3m})$ and $\mathcal{A}_{j+1+m}^\e\sim N^{m+1}\e^2\e^{3m}\sim\e^m$). For details, see Chapter 10 Part III in\cite{GSRT13} or Section 3.1 in \cite{PSS14}.

First, we report the main result achieved in \cite{GSRT13}:
\begin{theorem}[Theorem 5 in \cite{GSRT13}]\label{thm:GSRT}
Assume the repulsive potential $\Phi$ satisfies the following assumptions:
\begin{itemize}
\item[(i)] $\Phi:\R^3\to\R$ is a radial, nonnegative, non increasing function supported in the unit ball of $\R^3$, of class $\mathcal{C}^2$ in $\{x\in\R^3,\ 0<|x|<1\}$, unbounded near zero, approaches zero as $|x|\to1^-$ with bounded derivatives, and $\nabla\Phi$ vanishes only on $|x|=1$;
\item[(ii)]\ for $|x|\in(0,1)$, 
\be\label{eq:potential-cond}
|x|\Phi''(|x|)+2\Phi'(|x|)\geq 0\,.
\ee
\end{itemize}
Let $f(0):\R^3\times\R^3\to\R_+$ be a continuous density of probability such that for  $\beta>0$ $$\|f(0)\exp{[\frac{\beta}{2}|v|^2]}\|_{L^\infty}<\infty.$$ 
Consider the system of $N$ particles, initially distributed according to $f_0$ and asymptotically independent, governed by Eq. \eqref{eq:N}. Then, in the BG-limit,  its distribution function converges to the solution of the  Boltzmann equation \eqref{eq:BE} with a bounded cross-section, depending on $\Phi$ implicitly, and with initial data $f_0$, in the sense of observables, for short times.  
\end{theorem}

\begin{remark}\label{rem:convergence}
The convergence established by Theorem~\ref{thm:GSRT} is ``in the sense of observables", that means convergence uniformly in $t$ and $x$, after testing against a compactly supported function of $v$. Precisely, we say that $f_j^N$ converges to $f_j$ in the sense of the observables if, for any $\varphi\in\mathcal{C}_c^0(\R^{3j})$, \[ \int \varphi(v_1,\dots,v_j)f_j^N(z_1,\dots,z_N)\,dv_1\dots dv_j \to \int \varphi(v_1,\dots,v_j)f_j(z_1,\dots,z_N)\,dv_1\dots dv_j.\]
\end{remark}

\begin{remark}
Item $(ii)$ in Theorem~\ref{thm:GSRT} is a technical assumption due to the strategy adopted in the proof. Indeed, the authors need the scattering angle to be invertible in the impact parameter variable. Condition \eqref{eq:potential-cond} ensures that the scattering angle is a monotone function of the impact parameter, and hence invertible (see also Appendix in \cite{PSS14} for a detailed explanation). 
\end{remark}

From a physical point of view, the assumptions on the class of potentials for which the Boltzmann equation has been proved to hold is not satisfying, since it is heuristically expected to be valid independently of the details of the scattering. 

%The limitation in Theorem \ref{thm:GSRT} has been overcome in \cite{PSS14}, where a different approach allows to consider all potentials which are short range and stable in the sense of statistical mechanics\footnote{We say that a two-body potential $\Phi:\R^3\to\R_+$ is {\it stable} if there exists a positive constant $B$, such that $$\sum_{i=1}^N\sum_{\substack{j=1\\j < i}}^N \Phi(x_i-x_j)\geq -BN.$$}(see \cite{Ruelle}). In particular, singularities in the differential cross-section are allowed. Before stating the result in \cite{PSS14} we need to modify the functional spaces given in Def.~\ref{def:fspace} and Def.~\ref{def:fjspace} according to the following definitions:
%%
%\begin{definition}\label{def:Y-j}
%Let $Y_{j,\beta}$ be the space of Borel functions on $\mathcal{M}_{j}$ such that
%\[ \|f_j\|_{j,\beta}=\sup_{(z_1,\dots,z_j)\in\mathcal{M}_j} |f_j(z_1,\dots,z_j)|(\beta/2\pi)^{-\frac{3}{2}j}e^{\beta H(z_1,\dots,z_j)} <\infty\,, \]
%where $H(z_1,\dots,z_j)=\sum_{i=1}^j\frac{v_i^2}{2}+\frac{1}{2}\sum_{\substack{i,k=1\\k\neq i}}^j\Phi\left(\frac{x_i-x_k}{\e}\right)$.
%\end{definition} 
%%
%\begin{definition}\label{def:Y-b}
%For all $b>0$, $Y_{b,\beta}$ is the space of sequences of functions $f=\{f_j\}_{j\geq 1}$ such that the following norm is finite 
%%
%\[  \|f\|_{b,\beta}=\sup_{j}\ b^{-j}\|f_j\|_{j,\beta}\,.  \]
%% 
%\end{definition}
%%
We are now ready to state the following

\begin{theorem}[Theorem 1 in \cite{PSS14}]\label{thm:PSS}
Consider a two-body radial potential $\Phi:\R^3\to\R$ supported in $|q|<1$ and non increasing in $|q|$. We assume 
\begin{itemize}
\item[(i)]\ either $\Phi\in\mathcal{C}^2(\R^3)$, or $\Phi\in\mathcal{C}^2(\R^3\setminus\{0\})$ and  $\Phi(|x|)\to\infty$, as  $|x|\to 0$;
\item[(ii)]\ the initial data of the Boltzmann equation $f(0):\R^3\times\R^3\to\R_+$ is a probability density, continuous and such that,  for  $\beta>0$, $$\|f(0)\exp{[\frac{\beta}{2}|v|^2]}\|_{L^\infty(\R^3\times\R^3)}<\infty\,.$$ 
\item[(iii)]\, \ \ for any $N$, $W^N(0)$ is a probability density on the phase space $\mathcal{M}_N$, symmetric in the exchange of particles, with reduced marginals $\{\tilde{f}_j^N(0)\}_{\{j=1\}}^N$ such that 
$\|\tilde{f}_j^N(0)\|_{j,\beta}<e^{bj}$,
for $b,\beta>0$ and given $\tilde{f}_j^N(0)$ and $f_j(0)=f^{\otimes j}(0)$, we assume
\[\lim_{\e\to 0}\tilde{f}_j^N(0)=f_j(0)\,,\]
in the BG-limit, uniformly on compact sets in $\mathcal{M}_j$.
\end{itemize}
Then, there exists  $t_0>0$ such that, $\forall t<t_0$ and $\forall j\in\mathbb{N}$, $\tilde{f}_j^N(t)$ and $f_j(t)=f(t)^{\otimes j}$ exist and 
\[\lim_{\e\to 0}\tilde{f}_j^N(t)=f_j(t)\,,\]
in the BG-limit, uniformly on compact sets in $\Omega_j=\{(z_1,\dots,z_j)\in\mathcal{M}_j\,:\,(x_i-x_k)\wedge(v_i-v_k)\neq 0\}$, with $f(t)$ solution to \eqref{eq:BE} with initial datum $f(0)$.
\end{theorem}
The key ingredient here is to consider the formulation \eqref{eq:BE-HS} for the collision operator, which does not require the inverse of the scattering angle to exist as a single-valued function. Roughly speaking, the problem in considering the formulation \eqref{eq:Q} is to invert the map
$\bm\nu\to\omega\,.$
This is a big technical and conceptual difference with respect to the hard-sphere case, in which $\omega=\bm\nu$. 
\medskip

{\bf Sketch of the proof of Theorem~\ref{thm:PSS}.}
Following Lanford's proof, we want to compare $\tilde{f}_j^N$ and $f_j$. Step $(a)$ is achieved exactly as in Sect.~\ref{sect:HS}, according to the new definition \eqref{eq:reduced-marg} and the formulation \eqref{eq:BBGKY-shortpot}. Obviously, the time restriction in Theorem \ref{thm:PSS} is a consequence of step $(a)$. As for step $(b)$, we have to compare the IBF and the BBF. In particular, we want to show that, even for smooth short-range potentials, the sets which lead to a dynamics which is not close to the one of the Boltzmann flow are negligible in the BG-limit.   

In the case of short-range potentials, the IBF differs from the BBF because:
\begin{itemize}
\item collisions occur at distance $\e$;
\item recollisions may occur;
\item the scattering is not instantaneous;
\item multiple collisions may occur.
\end{itemize}
As consequence of the third point, we have to carefully analyse the low energy collisions, the high energy collisions, the central collisions and the recollisions. In particular, a dramatic difference may occur if: a particle created in the IBF interacts for long time with its progenitor; a couple of particles in the IBF undergoes a recollision. We study each event separately:
\medskip

{\it (a) A particle in the IBF interacts for long time with its progenitor.} This issue was not present in the hard-sphere case. It is overcome by cutting-off the impact vectors and the velocities  $(\bm\nu_i, v_{j+1})$ leading to the singular scattering and by showing that 
the contribution they give to the integrals is small in the BG-limit. To prove that, we need to estimate the scattering time $t_*$ . As it is shown in Lemma~1 in \cite{PSS14}, it can be bounded as follows:
\be\label{eq:scattering-time}
t_*\leq\frac{A}{\rho V}\e\,,
\ee
where $A$ is an absolute constant, $\rho$ is the impact parameter and $V$ is the relative velocity before the scattering takes place. Hence the scattering time may be too long if the relative velocity involved in the bound \eqref{eq:scattering-time} is small or if the impact parameter $\rho$ is close to zero (i.~e. a central collision occurs). To avoid these pathologies, we cut off small relative velocities and the parameters $(\nu_i, v_{j+1})$ leading to a central collision. The contribution given to the integrals by the set of cut off variables is negligible in the BG-limit. 
\medskip

{\it (b) A particle has a very large velocity.} 
This occurence is present in the hard-sphere case too and it is controlled in the same way, by cutting-off the large values of $|(v_{j+1},\dots,v_{j+n})|$. The integral over the cut-off region is small because $f_{j+n}^N\in X_{j+n,\beta}$.
\medskip

{\it (c) A couple of particles in the IBF undergoes a recollision.}
This is the most delicate task, because concentrations of measure in the differential cross-section may occur so that the integral over negligible sets can give a contribution of positive measure. Here we just give an idea of the main issues and we refer to \cite{PSS14} Sect.~7.2 for a detailed description of the technical part. We need to demonstrate that the contribution of recolliding trajectories is negligible in the limit $\e\to 0$.
To do that, the strategy adopted in \cite{PSS14} is based on three main ideas:
$(i)$ to work on the BBF instead of looking at the IBF and to exploit its simpler structure; $(ii)$ to perform the integrals on the time variables; $(iii)$ to keep using $\bm\nu$ instead of switching to $\omega$. Because of this latter point, the Boltzmann collision operator emerges in the form \eqref{eq:BE-HS} rather than in the usual formulation given by \eqref{eq:Q}.\\
First, we define by words the set 
\[\mathcal{N}(\delta):=\{\mbox{a couple of particles in the BBF is getting closer than } \delta>\e \}\,,  \]
where $\delta$ is chosen as a function of $\e$, vanishing as $\e\to 0$. 
We observe that 
\[\lim_{\delta\to 0}\bm{1}_{\mathcal{N}(\delta)}=\bm{1}_{\mathcal{N}}\,,\]
with $\mathcal{N}:=\{\mbox{couples of particles in the BBF which recollide pointwise} \}$ and
 $\mathcal{N}$ is a zero-measure set with respect to the measure $d\Lambda$. This is shown by making use of the time integrals $\int_0^t dt_1\int_0^{t_1}dt_2\dots\int_0^{t_{n-1}}dt_n$.
Hence, we are left with the control of the contribution given by the complement of the set $\mathcal{N}(\delta)$, defined as $\mathcal{N}(\delta)^c$. We notice that in $\mathcal{N}(\delta)^c$ the BBF is close to the IBF when the scattering times are small and the velocities are not large (we already cut off the long scattering times in  $(a)$ and large velocities in $(b)$).
  \qed
  \medskip
 
Following the sketch of proof above, it is possible to extend Theorem~\ref{thm:PSS} to stable short-range potentials:

\begin{theorem}[Theorem 1' in \cite{PSS14}]
Let $\Phi(q)$ be a stable radial potential, with support $|q|<1$. 
Under the Hypotheses $(i)-(iii)$ in Theorem~\ref{thm:PSS},
there exists $t_0 >0$ such that, for any positive $t<t_0$ and $j\in \N$,
the series expansions are absolutely convergent (uniformly in $\e$),
and
\be
\lim_{\substack{\e\rightarrow 0\\ N\e^2=1}} f_j^N(t) = f_j(t)
\ee
uniformly on compact sets in $\Omega_j.$
\end{theorem} 

\begin{remark}
The lack of explicit estimates in the proof of the above Theorem is due to the difficulty in reproducing a bound of type \eqref{eq:scattering-time} in the case of stable potentials, due to the possible presence of trapping orbits in the attractive region. 
\end{remark}
 
Under the assumptions of Theorem \ref{thm:PSS} on the potential and further assumptions on the initial data, it is possible to compute explicitly the rate of convergence:
\begin{theorem}[Theorem 2 in \cite{PSS14}]
Assume the hypotheses of Theorem~\ref{thm:PSS} to hold. Moreover, assume that the potential $\Phi$ is non-increasing and that :
\[
\begin{split}
&\sup_{|x_i-x_k|>\e}e^{\beta\sum_{i=1}^j\frac{v_i^2}{2}}|f_j^N(0)-f(0)^{\otimes j}|\leq C^j\e\,,\\
& e^{\beta\frac{v^2}{2}}|f(0,x,v)-f(0,x',v)|\leq L|x-x'|\,, \mbox{ for some } L\geq 0\,.
\end{split}
\]
Then, for $t\in[0,t_0)$, $(z_1,\dots,z_j)\in\Omega_j$, there exist constants $C$, $\gamma>0$, such that for any $j\geq 1$ and $\e$ small enough, the following estimates hold
\[|f_j^N(t,z_1,\dots,z_j)-f(t)^{\otimes j}(z_1,\dots,z_j)|\leq C^j\e^\gamma\,,\ \ \gamma<\frac{1}{6}\,.\]
 \end{theorem}

%%%%%%%%%%%%%%%%%%%%%%%%%%%%%%%%%%%%%%%%%%%%%
%%%%%%%%%%%%%%%%%%%%%%%%%%%%%%%%%%%%%%%%%%%%%

\section{Beyond the short range} 
\label{sect:long}

Apart from the long time validity, the other interesting and natural open question concerning the derivation of the Boltzmann equation is whether the results \cite{La75,GSRT13,PSS14} can be extended to the case of  long-range interactions. From a phenomenological point of view, it should be possible to show that the Boltzmann equation emerges from the microscopic classical dynamics, at least for potentials of the form $\Phi(|x|)=\frac{1}{|x|^\alpha}$, for an appropriate choice of $\alpha$. Heuristically,  this was justified by Maxwell in his paper \cite{Ma67}, where he proposed Eq.~\eqref{eq:BE} (tested against a smooth function of the velocity variable) to be a  good approximation of the dynamics of a rarefied gas with intermolecular force an inverse power law potential. The question here is to make rigorous Maxwell's argument, for a reasonable class of long-range potentials. It has been investigated in \cite{DP99} in the simpler linear case.

In this section we want to give an overlook on this open problem, underlining the difficulties one has to face.

The first obstacle one has to cope with is to define the scaling limit. Indeed, in the case of hard-core and short-range potentials, the scale parameter $\e$, which goes to zero, represents the diameter of particles or the range of the interaction, respectively. Because of the long tail of the potential,  $\e$ cannot represent the range of interaction anymore. Therefore, a revised version of the BG-limit seems to be necessary to state the problem in a rigorous mathematical way, taking into account the mean-field effects appearing at large distances in the long-range interaction.    

The second difficulty is to show the well-posedness for the BBGKY hierarchy. In fact, consider the particular case of an interaction given by an hard-sphere dynamics plus a long tail $\Phi$ for $|x|>\e$. Under the usual hyperbolic scaling of space and time \eqref{eq:scaling}, a new term related to the long tail of the potential appears in the hierarchy: 
\be\label{eq:BBGKY-long}
\begin{split}
(\partial_t+\sum_{i=1}^j v_i\cdot\nabla_{x_i}){f}_j^N&=\tilde{\mathcal{L}}_{j}^\e{f}_j^N + \tilde{\mathcal{C}}^\e_{j+1}{f}_{j+1}^N+\mathcal{L}^\e_jf^N_j\\
&+\frac{N-j}{\e}\sum_{i=1}^j\int dx_{j+1}\int dv_{j+1}\,\nabla_{x_i}\Phi\left(\frac{x_i-x_{j+1}}{\e}\right)\cdot\nabla_{v_i}{f}^N_{j+1}\,,
\end{split}
\ee
where we used the notations introduced in Sect.~\ref{sect:HS}. The difficulty here is to get a priori estimates on the derivative with respect to $v_i$ of the reduced marginal ${f}^N_{j+1}$. One could use the ideas proposed by Maxwell in his heuristic presentation of the Boltzmann equation, based on the convergence in the sense of observables (as defined in Remark~\ref{rem:convergence}), i.e. the weak formulation may help to give sense to the third term on the r.h.s. of Eq. \eqref{eq:BBGKY-long}.

The situation becomes even more problematic when looking at the Coulomb potential $\Phi(|x|)=\frac{1}{|x|}$ ($\alpha=1$). In this case the collision integral in the r.h.s. of the Boltzmann equation makes no sense whatever choice of $f$. This suggests to replace the Boltzmann equation by a different model. Indeed,  the slow decay at infinity of the potential makes the so-called grazing collisions to be of leading importance in the macroscopic behaviour of the gas. This problem was pointed out in 1936 by Landau \cite{Landau}, who proposed a modified equation to describe the effect of grazing collisions. The Landau equation reads
\be\label{eq:Landau}
\begin{split}
(\partial_t+v\cdot\nabla_{x})&f(t,x,v)\\
=& \nabla_{v}\cdot\int_{\R^3} dv_* \frac{P_{(v-v_*)^{\perp}}}{|v-v_*|}\{f(t,x,v_*)\nabla_{v}f(t,x,v)-f(t,x,v)\nabla_{v_*}f(t,x,v_*)\}\,.
\end{split}
\ee
where $P_{(w)^\perp}$ is the orthogonal projection on the subspace orthogonal to $w\in\R^3$.

Up to now, there are only very few mathematical results about the Landau model \eqref{eq:Landau}. The validity problem and the well-posedness of the 
equation are open questions of primary interest and importance, especially so because of the several applications involving the 
Landau equation. The Cauchy problem associated to the homogeneous Landau equation has been studied in  \cite{Villani, Desvillettes}, where weak solutions are proven to exist. Uniqueness is proved in \cite{Fournier} once the solution is known to belong to $L^\infty$. In the non-homogeneous case, the only available result is due to Guo  \cite{Guo}, who proved that there exists a global unique classical solution of \eqref{eq:Landau} for small perturbations of the equilibrium.

As for the validity problem, the only attempt to derive the Landau equation from the Hamiltonian dynamics is contained in \cite{BPS13}, where a consistency result is achieved, in the weak-coupling limit starting from a system of N particles interacting via a rescaled smooth short-range potential.
In this regime, a given particle undergoes a huge number of collisions in the kinetic time, but the two-body potential is weakened, and hence the variance of the total momentum variation remains finite. The key idea in \cite{BPS13} is based on the fact that $f_j^N$ cannot be smooth. 
If it were, we would have a trivial free dynamics. Hence we make the ansatz
 $$f_j^N=g_j^N+\gamma_j^N$$
where $g_j^N$ is smooth and $\gamma_j^N$ is strongly oscillating. This allows to find a system of coupled equations for $g_j^N$ and $\gamma_j^N$. In particular, the equation for $\gamma_j^N$ can be solved in terms of $g_j^N$. This leads to the following hierarchy:
\be\label{eq:Landau-hier}
\begin{split}
g_j^N(t)&= \mathcal{S}(t)f_j(0)+\frac{N-j}{\sqrt\e}\int_0^t d\tau\,\mathcal{S}(t-\tau)\,\mathcal{C}_{j+1}^\e g_{j+1}^N (\tau)\\
  &+\frac {N-j}{ \e}\int _0^t  d\tau\,\int_0^\tau d\sigma\,\mathcal{S}(t-\tau)\,\mathcal{C}^\e _{j+1}\,\mathcal{U}_{j+1}^\e (\tau-\sigma )\,\mathcal{T}_{j+1}^\e\, g_{j+1}^N(\sigma)\,,
\end{split}
\ee
where $\mathcal{S}$ is the generator of the free-flow, $\mathcal{U}_{j+1}^\e$ is the generator of the evolution of $\gamma_j^N$, $\mathcal{C}_{j+1}^\e$ is a collision operator and  $\mathcal{T}_{j+1}^\e$ is the Liouville operator restricted to $j+1$ particles.
By perturbing \eqref{eq:Landau-hier} up to the second order in time, we obtain $g_j^N(t)\to f(t)^{\otimes j}$, where $f(t)$ is a solution to the Landau equation, and $\gamma_j^N\to 0$, where the convergence has to be understood in distributional sense. 

This result is not fully satisfactory because of the lack of control of higher order terms.
Therefore, the rigorous mathematical validity of the Landau equation is an open problem, even for small time intervals.  

%%%%%%%%%%%%%%%%%%%%%%%%%%%%%%%%%%%%%%%%%%%%%
%%%%%%%%%%%%%%%%%%%%%%%%%%%%%%%%%%%%%%%%%%%%

%
\begin{acknowledgement}
The author is supported by the fund ``Forschungskredit UZH FK-15-108".
\end{acknowledgement}

\end{document}